# Implementing Multiple Modes of the Perpendicular Magnetization Switching within a Single Spin Orbit Torque Device

Tongxi Liu, Zhaohao Wang*, *Senior Member, IEEE,* Min Wang, Chao Wang, Bi Wu, Weiqiang Liu, *Senior Member, IEEE,* and Weisheng Zhao*, *Fellow, IEEE*.

*Abstract*—Spin orbit torque (SOT) has been considered as one of the promising technologies for the next-generation magnetic random access memory (MRAM). So far, SOT has been widely utilized for inducing various modes of magnetization switching. However, it is challenging to integrate multiple modes of magnetization switching together. In this work we propose a method for implementing both unipolar and bipolar switching of the perpendicular magnetization within a single SOT device. The mode of switching could be easily altered by tuning the amplitude of the applied current. We show that the field-like torque plays an important role in the switching process. The field-like torque induces the precession of the magnetization in the case of unipolar switching, whereas it helps to generate an effective z-component torque in the case of bipolar switching. In addition, the influence of key parameters on the mode of switching is discussed. Our proposal could be used to design novel reconfigurable logic circuits in the near future.

## I. Introduction

Magnetic random access memory (MRAM) has become a promising candidate for both embedded and standalone applications [1], thanks to its non-volatility, low power, high speed, and nearly unlimited endurance. To date, spin-transfer torque (STT) MRAM has made great progress in both academia and industry. Nevertheless, new-generation MRAM has always been explored by researchers to outperform the STT-MRAM. Typically, spin orbit torque (SOT) MRAM shows great potential in non-volatile memory and in-memory computing [2]-[4]. Compared with the STT-MRAM, the read and write paths are separated in the SOT-MRAM, resulting in higher reliability. Furthermore, SOT-driven magnetization switching is as fast as several hundreds of picoseconds [5]-[6], which qualifies the SOT-MRAM to be used in high-level caches.

Actually, the SOT-driven magnetization switching is achieved under the joint effects of multiple factors, such as magnetic anisotropy field, Gilbert damping torque, field-like torque, etc. More notably, by changing the relative proportions of these factors, the behavior of magnetization switching varies dramatically. Overall, both unipolar and bipolar switching have been proposed with different SOT mechanisms [7]-[13]. For the unipolar switching, the magnetization is switched to the opposite state once the SOT current is applied. For the bipolar switching, the magnetization is switched to a definite state depending on the direction or path of the applied SOT current. The combination of unipolar and bipolar switching could benefit the function extension of the SOT-based memories or circuits. However, up to now these two modes of magnetization switching are implemented with different devices separately, which degrades the design flexibility of the related circuits.

In this letter, we propose a novel scheme that implements both unipolar and bipolar switching of the perpendicular magnetization within a single SOT device. The mode of switching is only dependent on the amplitude of the applied current. The change of switching mode is mainly attributed to the modulation of the field-like torque. Our proposal makes it possible to design the SOT-based memories or circuits with good reconfigurability.

## II. The Device Model

In this study, the magnetization switching occurs in a common SOT magnetic tunnel junction (MTJ) with perpendicular anisotropy, as illustrated in Fig. 1. No special structure is required in this device. A charge current passing through the heavy metal induces the SOT which switches the perpendicular magnetization of the free layer (FL). Generally an additional bias field is applied to break the symmetry so that the switching process becomes deterministic. This bias field could be generated inside the device by using antiferromagnet [14]-[15] or magnetic hard mask [16]. The magnetization dynamics of the FL can be described by a modified Landau–Lifshitz–Gilbert (LLG) equation, as follows:

This work was supported by the Beijing Municipal Science and Technology Project under Grant Z201100004220002, the Fundamental Research Funds for the Central Universities under Grant YWF-21-BJ-J-1043, and the National Natural Science Foundation of China under Grant 61704005. (Corresponding authors: Zhaohao Wang and Weisheng Zhao).

T. Liu, Z. Wang, M. Wang, C. Wang and W. Zhao are with Fert Beijing Institute, MIIT Key Laboratory of Spintronics, School of Integrated Circuit Science and Engineering, Beihang University, Beijing 100191, China.
B. Wu and W. Liu are with the College of Integrated Circuits, Nanjing University of Aeronautics and Astronautics, Nanjing 211106, China.

TABLE I
MAGNETIC PARAMETERS

| Parameters | Value |
| --- | --- |
| Free Layer Thickness | 1 nm |
| Damping Constant ($\alpha$) | 0.05 |
| Spin Hall Angle | 0.3 |
| Saturation Magnetization | $1 \times 10^6$ A/m |
| Effective Anisotropy Constant | $1.5 \times 10^5$ J/m$^3$ |

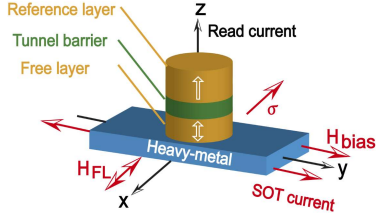

Fig. 1. The device structure and coordinate system in this work.

$$\frac{\partial \mathbf{m}}{\partial t} = -\gamma \mu_0 \mathbf{m} \times \mathbf{H}_{eff} + \alpha \mathbf{m} \times \frac{\partial \mathbf{m}}{\partial t}$$
$$- \lambda_{DL} \xi J \mathbf{m} \times (\mathbf{m} \times \boldsymbol{\sigma}) - \lambda_{FL} \xi J \mathbf{m} \times \boldsymbol{\sigma} \quad (1)$$

where $\mathbf{m}$ and $\boldsymbol{\sigma}$ are the unit vectors of the FL magnetization and SOT-induced spin polarization, respectively. $J$ is the SOT current density. $\lambda_{DL}$ and $\lambda_{FL}$ represent the strengths of the damping-like and field-like torques, respectively. $\lambda_{DL}$ is equivalent to the spin Hall angle. Other variables are described elsewhere [10]. Some parameters are configured as Table I.

### III. RESULTS AND DISCUSSIONS

In this work the field-like torque must be strong enough to implement multiple modes of switching within the above magnetic device. Strong field-like torque has been reported in previous works [17]. Here $\lambda_{FL}/\lambda_{DL} = 4$ is chosen for a preliminary study. Accordingly, macrospin simulation results under the various current densities are shown in Fig. 2, where unipolar and bipolar switching could be clearly observed. First, no switching occurs when the current density is insufficient, since the torque is too weak to overcome the energy barrier. Second, unipolar switching is achieved if the current density is set to an intermediate value (see Fig. 2(b)(e)). Finally, the switching process becomes bipolar while the current density is further increased (see Fig. 2(c)(f)). Therefore, the mode of switching could be easily changed by adjusting the current density. The detailed mechanisms are analyzed below.

The large field-like torque plays a dominate role in the unipolar switching process. According to (1), the field-like torque is equivalently induced by an in-plane magnetic field ($\mathbf{H}_{FL}$). For $\lambda_{FL}/\lambda_{DL} = 4$ and $J = 6 \times 10^{11}$ A/m$^2$, this equivalent magnetic field is around $\lambda_{FL}\xi J/\gamma \approx 237$ mT, which is much larger than the bias field (20 mT). In this case, the magnetization vector almost precesses around the $\mathbf{H}_{FL}$ with a speed of $\gamma \mu_0 |\mathbf{H}_{FL}|$ [12], as shown in Fig. 3(a)(b). Thus, the magnetization vector will reach the in-plane direction (i.e. $m_z = 0$) after a delay of about $\pi/(2\gamma\mu_0|\mathbf{H}_{FL}|)$, which is in agreement with the results of Fig. 2(b)(e). Finally, the magnetization vector is stabilized at an equilibrium position under the action of various torques.

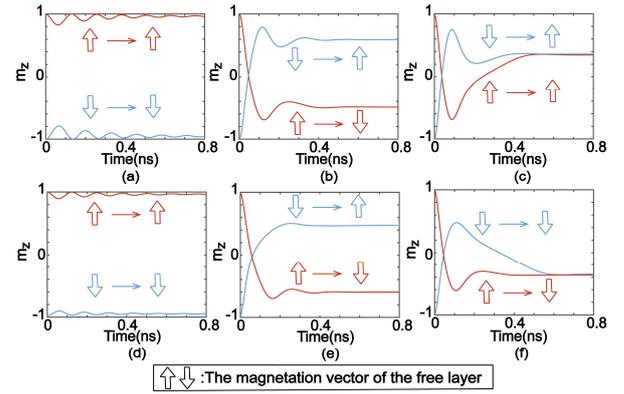

Fig. 2. Macrospin simulation results of z-component magnetization ($m_z$). (a)(d) no switching occurs while $J = 2 \times 10^{11}$ A/m$^2$. (b)(e) Unipolar switching occurs while $J = 6 \times 10^{11}$ A/m$^2$. (c)(f) Bipolar switching occurs while $J = 7 \times 10^{11}$ A/m$^2$. (a)(b)(c) the current is applied along +y axis. (d)(e)(f) the current is applied along –y axis.

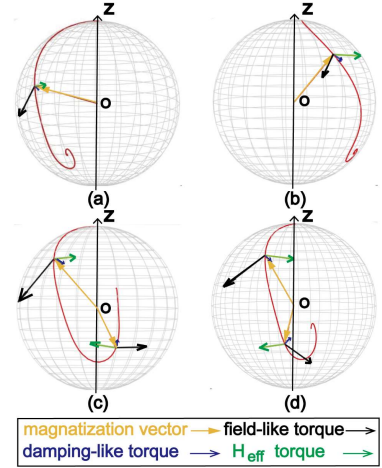

Fig. 3. The trajectories of magnetization vector and key torques during the switching process. (a)(b) Unipolar switching. (c)(d) Bipolar switching. Here other torques are not shown for the clarity. It is seen from (c) and (d) that the $\mathbf{H}_{eff}$ torque has +z and –z components, respectively, which are mainly contributed by the bias field ($\mathbf{H}_{bias}$).

The bipolar switching occurs if a larger current density is applied. In this case, the field-like torque is enhanced so that the magnetization vector is driven closer to the axis of $\boldsymbol{\sigma}$. Then the torque induced by the bias field ($\mathbf{H}_{bias}$) is nearly aligned to $\pm\boldsymbol{\sigma} \times \mathbf{H}_{bias}$. Note that $\boldsymbol{\sigma}$ and $\mathbf{H}_{bias}$ are parallel to x and y-axes, respectively (see Fig. 1), thus this torque is almost oriented towards $\mp z$ axis. Depending on the direction of the applied current, the magnetization vector is switched to the +z or -z axis, as shown in Fig. 3(c)(d). Overall, the joint effects of the bias field and huge field-like torque lead to the bipolar switching.

It is important to mention that the role of the bias field becomes significant only when the current density is sufficiently large. This conclusion could be explained by Fig. 4(a), where a non-zero bias field induces an effective torque to pull the magnetization vector back, leading to bipolar switching. In contrast, the impact of the bias field is negligible in the case of unipolar switching, as shown in Fig. 4(b). The difference between Fig. 4(a) and (b) is attributed to the various x-component magnetization ($m_x$). Specifically, since $\mathbf{H}_{bias}$ is aligned to +y-axis, the z-component effective torque is contributed by $\pm\gamma\mu_0 \mathbf{m}_x \times \mathbf{H}_{bias}$. Stronger current density leads to larger $m_x$ (see Fig. 4(a)), and hence more easily drives the magnetization vector towards z-axis.

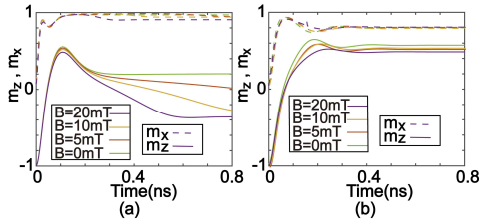

Fig. 4. The influence of the bias field on the magnetization switching. (a) Bipolar switching. (b) Unipolar switching.

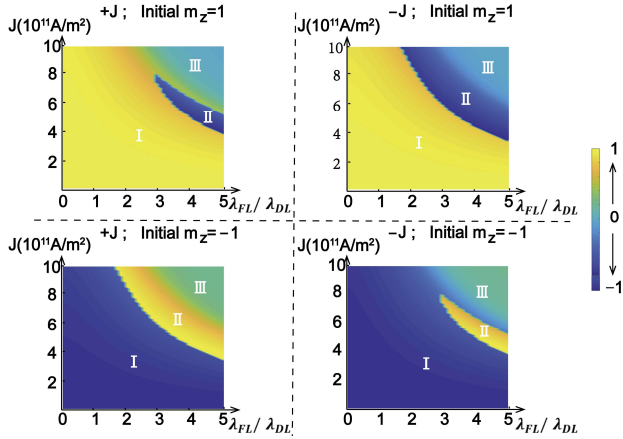

Fig. 5. The phase diagram of $m_z$ as a function of the $\lambda_{FL}/\lambda_{DL}$ and $J$. Regions I, II, and III indicate, respectively, no switching, unipolar switching, and bipolar switching.

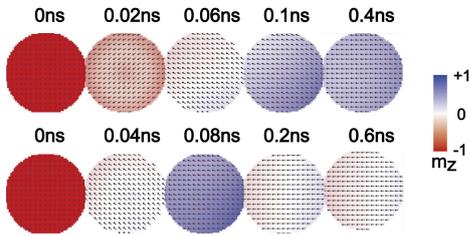

Fig. 6. Typical simulation results of micromagnetic configuration during the switching process. Here the DMI magnitude is 0.3 mJ/m$^2$.

The influence of key parameters on the switching mode is further discussed. Fig. 5 shows the phase diagram of the final-state $m_z$ as a function of the $\lambda_{FL}/\lambda_{DL}$ and $J$. Overall, the bipolar switching occurs when both the current density and the field-like torque are set to be large enough. In this case, the combination of stronger field-like torque and larger current density causes magnetization vector to approach in-plane direction, favoring the bias-field-induced torque, as illustrated by Fig. 3(c)(d). On the other hand, the unipolar switching is obtained in a relatively narrow range of parameters, where the field-like torque suppresses the other factors and leads to the precession of the magnetization vector, as shown in Fig. 3(a)(b).

The proposed mechanism of the magnetization switching is also validated through micromagnetic simulation. Here the Dzyaloshinskii-Moriya interaction (DMI) is taken into account. Typical results are shown in Fig. 6, where the magnetization switching is completed through magnetic domain dynamics. Both the unipolar and bipolar switching are demonstrated with the appropriate parameter settings. However, the DMI has a dramatic influence on the characteristic of the magnetization switching. The unipolar switching disappears in the case of strong DMI (not shown here). Nevertheless, we found that the unipolar switching could be recovered by increasing the exchange constant or decreasing the bias field. The reason may be that the uniformity of the magnetization are enhanced.

## IV. CONCLUSION

We have proposed a novel switching scheme for the perpendicular-anisotropy SOT device. The switching mode of the device could be converted between the unipolar and bipolar types, only by tuning the amplitude of the current density. For the unipolar switching, the strong field-like torque mainly governs the magnetization dynamics. For the bipolar switching, the bias field and field-like torque jointly determine the polarity of the magnetization switching. Our proposal could breed a multifunction SOT device, especially suitable for the design of the reconfigurable circuits.

## REFERENCES

[1] B. Dieny, *et al.*, "Opportunities and challenges for spintronics in the microelectronics industry," *Nature Electron.*, vol. 3, no. 8, pp. 446–459, 2020, doi: 10.1038/s41928-020-0461-5.

[2] Z. Guo, *et al.*, "Spintronics for Energy- Efficient Computing: An Overview and Outlook" *Proc. IEEE*, pp. 1-20, 2021, doi: 10.1109/JPROC.2021.3084997.

[3] S.-W. Lee, and K.-J. Lee, "Emerging three-terminal magnetic memory devices," *Proc. IEEE*, vol. 104, no. 10, pp. 1831–1843, Oct. 2016. doi: 10.1109/JPROC.2016.2543782.

[4] Z. I. Chowdhury, *et al.*, "Spintronic In-Memory Pattern Matching," *IEEE J. Explor. Solid-State Computat. Devices Circuits,* vol. 5, no. 2, pp. 206–214, Dec. 2019, doi: 10.1109/JXCDC.2019.2951157.

[5] H. Honjo, *et al.*, "First demonstration of field-free SOT-MRAM with 0.35 ns write speed and 70 thermal stability under 400∘C thermal tolerance by canted SOT structure and its advanced patterning/SOT channel technology," *IEDM Tech. Dig., San Francisco, CA, USA*, Dec. 2019, p. 28, doi: 10.1109/IEDM19573.2019.8993443.

[6] W. Cai, *et al.*, "Sub-ns field-free switching in perpendicular magnetic tunnel junctions by the interplay of spin transfer and orbit torques," *IEEE Electron Device Lett.*, vol. 42, no. 5, pp. 704–707, May 2021, doi: 10.1109/LED.2021.3069391.

[7] S. Fukami, T. Anekawa, C. Zhang, and H. Ohno, "A spin-orbit torque switching scheme with collinear magnetic easy axis and current configuration," *Nature Nanotechnol.*, vol. 11, pp. 621–625, 2016, doi: 10.1038/nnano.2016.29.

[8] M. Wang, *et al.*, "Field-free switching of a perpendicular magnetic tunnel junction through the interplay of spin-orbit and spin-transfer torques," *Nature Electron.*, vol. 1, no. 11, pp. 582–588, 2018, doi:10.1038/s41928-018-0160-7.

[9] M. Yang, J. Luo, Y. Ji, H.-Z. Zheng, K. Wang, Y. Deng, Z. Wu, K. Cai, K. W. Edmonds, Y. Li, Y. Sheng, S. Wang, and Y. Cui, "Spin logic devices via electric field controlled magnetization reversal by spinorbit torque," *IEEE Electron Device Lett.*, vol. 40, no. 9, pp. 1554–1557, Sep. 2019, doi: 10.1109/led.2019.2932479.

[10] M. Wang, Z. Wang, C. Wang, and W. Zhao, "Field-Free Deterministic Magnetization Switching Induced by Interlaced Spin−Orbit Torques," *ACS Appl. Mater. Interfaces,* vol. 13, no. 17, pp. 20763-20769, 2021, doi: 10.1021/acsami.0c23127.

[11] B. Chen, J. Lourembam, S. Goolaup, and S. T. Lim, "Field-free spin-orbit torque switching of a perpendicular ferromagnet with dzyaloshinskii-moriya interaction," *Appl. Phys. Lett.*, vol. 114, no. 2, Jan. 2019, Art. no. 022401, doi: 10.1063/1.5052194.

[12] W. Legrand, R. Ramaswamy, R. Mishra, and H. Yang, "Coherent subnanosecond switching of perpendicular magnetization by the field-like spin-orbit torque without an external magnetic field," *Phys. Rev. Appl.*, vol. 3, no. 6, p. 064012, 2015,

doi: 10.1103/PhysRevApplied.3.064012.

[13] M. Wang, Z. Wang, X. Zhang, and W. Zhao, "Theoretical Conditions for Field-Free Magnetization Switching Induced by Spin-Orbit Torque and Dzyaloshinskii–Moriya Interaction," *IEEE Electron Device Lett.*, vol. 42, no. 2, pp. 148-151, Feb. 2021, doi: 10.1109/LED.2020.3043293.

[14] S. Fukami, C. Zhang, S. DuttaGupta, A. Kurenkov, and H. Ohno, "Magnetization switching by spin–orbit torque in an antiferromagnet—ferromagnet bilayer system," *Nature Mater.*, vol. 15, pp. 535–541, Feb. 2016, doi: 10.1038/nmat4566.

[15] Y.-W. Oh, S.-H. C. Baek, Y. M. Kim, H. Y. Lee, K.-D. Lee, C.-G. Yang, E.-S. Park, K.-S. Lee, K.-W. Kim, G. Go, J.-R. Jeong, B.-C. Min, H.-W. Lee, K.-J. Lee, and B.-G. Park, "Field-free switching of perpendicular magnetization through spin–orbit torque in antiferromagnet/ferromagnet/oxide structures," *Nature Nanotechnol.*, vol. 11, pp. 878–884, Jul. 2016, doi: 10.1038/NNANO.2016.109.

[16] K. Garello *et al.*, "Manufacturable 300 mm platform solution for fieldfree switching SOT-MRAM," *Proc. IEEE Symp. VLSI Technol.,* pp. T194–T195, Jun. 2019, doi: 10.23919/VLSIC.2019.8778100.

[17] A. Manchon, J. Železný, I. M. Miron, T. Jungwirth, J. Sinova, A. Thiaville, K. Garello, and P. Gambardella, ''Current-induced spin-orbit torques in ferromagnetic and antiferromagnetic systems,'' *Rev. Mod. Phys.*, vol. 91, no. 3, Sep. 2019, Art. no. 035004, doi: 10.1103/RevModPhys.91.035004.